\documentstyle[12pt]{article}
\begin{document}
\thispagestyle{empty}
\newcommand{\be}{\begin{equation}}
\newcommand{\ee}{\end{equation}}
\newcommand{\sect}[1]{\setcounter{equation}{0}\section{#1}}
\newcommand{\vs}[1]{\rule[- #1 mm]{0mm}{#1 mm}}
\newcommand{\hs}[1]{\hspace{#1mm}}
\newcommand{\mb}[1]{\hs{5}\mbox{#1}\hs{5}}
\newcommand{\bea}{\begin{eqnarray}}
\newcommand{\eea}{\end{eqnarray}}
\newcommand{\wt}[1]{\widetilde{#1}}
\newcommand{\und}[1]{\underline{#1}}
\newcommand{\ov}[1]{\overline{#1}}
\newcommand{\sm}[2]{\frac{\mbox{\footnotesize #1}\vs{-2}}
		   {\vs{-2}\mbox{\footnotesize #2}}}
\newcommand{\prt}{\partial}
\newcommand{\eps}{\epsilon}\newcommand{\p}[1]{(\ref{#1})}
\newcommand{\R}{\mbox{\rule{0.2mm}{2.8mm}\hspace{-1.5mm} R}}
\newcommand{\Z}{Z\hspace{-2mm}Z}
\newcommand{\cd}{{\cal D}}
\newcommand{\cg}{{\cal G}}
\newcommand{\ck}{{\cal K}}
\newcommand{\cw}{{\cal W}}
\newcommand{\vj}{\vec{J}}
\newcommand{\vl}{\vec{\lambda}}
\newcommand{\vz}{\vec{\sigma}}
\newcommand{\vt}{\vec{\tau}}
\newcommand{\vw}{\vec{W}}
\newcommand{\poiss}{\stackrel{\otimes}{,}}


\newcommand{\NP}[1]{Nucl.\ Phys.\ {\bf #1}}
\newcommand{\PLB}[1]{Phys.\ Lett.\ {B \bf #1}}
\newcommand{\PLA}[1]{Phys.\ Lett.\ {A \bf #1}}
\newcommand{\NC}[1]{Nuovo Cimento {\bf #1}}
\newcommand{\CMP}[1]{Commun.\ Math.\ Phys.\ {\bf #1}}
\newcommand{\PR}[1]{Phys.\ Rev.\ {\bf #1}}
\newcommand{\PRL}[1]{Phys.\ Rev.\ Lett.\ {\bf #1}}
\newcommand{\MPL}[1]{Mod.\ Phys.\ Lett.\ {\bf #1}}
\newcommand{\BLMS}[1]{Bull.\ London Math.\ Soc.\ {\bf #1}}
\newcommand{\IJMP}[1]{Int.\ J.\ Mod.\ Phys.\ {\bf #1}}
\newcommand{\JMP}[1]{Jour.\ Math.\ Phys.\ {\bf #1}}
\newcommand{\LMP}[1]{Lett.\ Math.\ Phys.\ {\bf #1}}

\renewcommand{\thefootnote}{\fnsymbol{footnote}}
\newpage
\setcounter{page}{0}
\pagestyle{empty}
\vs{12}
\begin{center}
{\LARGE {\bf N=4 Super NLS-mKdV Hierarchies }}\\[0.8cm]

\vs{10}
{\large E. Ivanov$^{a,1}$, S. Krivonos$^{a,2}$ and F. Toppan$^{b,3}$}
~\\
\quad \\
{\em {~$~^{(a)}$ JINR-Bogoliubov Laboratory of Theoretical Physics,}}\\
{\em 141980 Dubna, Moscow Region, Russia}~\quad\\
{\em ~$~^{(b)}$ Dipartimento di Fisica, Universit\`{a} di Padova,}\\
{\em Via Marzolo 8, 35131 Padova, Italy} 
\end{center}
\vs{6}

\centerline{ {\bf Abstract}}

\noindent $N=2$ extension of affine algebra $\widehat{sl(2)\oplus u(1)}$ 
possesses a hidden global $N=4$ supersymmetry and provides 
a second hamiltonian structure for a new $N=4$ supersymmetric 
integrable hierarchy defined on $N=2$ affine
supercurrents. This system is an $N=4$ extension  of at once 
two hierarchies, $N=2$ NLS and $N=2$ mKdV ones.    
It is related to $N=4$ KdV hierarchy via a generalized 
Sugawara-Feigin-Fuks construction which relates  
$N=2$ $\widehat{sl(2)\oplus u(1)}$ algebra to ``small'' $N=4$ SCA.
We also find the underlying affine hierarchy for 
another integrable system with the $N=4$ SCA second hamiltonian 
structure, ``quasi'' $N=4$ KdV hierarchy. It respects only 
$N=2$ supersymmetry. For both new hierarchies we 
construct scalar Lax formulations. We speculate that any  
$N=2$ affine algebra admitting a quaternionic structure 
possesses $N=4$ supersymmetry and so can be used to produce
$N=4$ supersymmetric hierarchies. This suggests a way of classifying 
all such hierarchies.

\vs{6}
\vfill
\rightline{March 1997}
\rightline{JINR E2-97-108}
\rightline{DFPD 97-TH-11}
\rightline{ hep-th/9703224}
{\em E-Mail:\\
1) eivanov@thsun1.jinr.dubna.su\\
2) krivonos@thsun1.jinr.dubna.su\\
3) toppan@mvxpd5.pd.infn.it}
\newpage
\pagestyle{plain}
\renewcommand{\thefootnote}{\arabic{footnote}}
\setcounter{footnote}{0}

\noindent{\bf 1. Introduction}. The integrable hierarchies 
of differential equations in $1+1$ dimensions
have been widely studied in the last several years both in the physical and 
in the mathematical contexts. There is 
an ever-growing 
evidence that their relevance should not be confined to the realm of
pure mathematics, but rather their beautiful mathematical structures 
show themselves naturally when investigating 
physical problems. At first they appeared in connection 
with the discretized versions,
via the matrix-models approach, 
of $2$-dimensional gravity (see \cite{dgz} and references therein). Recently
they sprang up rather unexpectedly in the 
domain of $4$-dimensional 
field theories as an underlying structure of the $N=2$
super Yang-Mills theories in the Seiberg-Witten approach \cite{sw}.  
The remarkable relation of the KdV-type hierarchies, 
via their second hamiltonian structure, to the conformal 
algebras (Virasoro algebra and its extensions) establishes a link
between such hierarchies and the string theories which has still to be
fully appreciated.

One is naturally led to investigate supersymmetric
extensions of the above bosonic hierarchies. Besides 
$N=1$ supersymmetric systems which were studied in 
a lot of papers starting from refs. \cite{mrk}, it is very interesting 
to consider the larger $N$ cases. 
One of the main sources of interest in extended supersymmetric 
hierarchies with $N > 1$ is that they could  
lead to a sort of ``unification or grand-unification'' of known 
hierarchies. It could happen that seemingly unrelated bosonic or 
lower supersymmetric ($N=1,2$) hierarchies are different 
manifestations of a single ``unifying'' larger $N$ supersymmetric 
hierarchy.

During the last years, $N=2$ supersymmetric hierarchies 
were a subject of many exhaustive studies. Much less was known 
about systems with higher $N$. In a series of papers  
[4-7] the first 
example of hierarchy with $N=4$ supersymmetry, $N=4$ super KdV system, 
was discovered and studied. It is related, through its second hamiltonian 
structure, to ``small'' $N=4$ superconformal algebra, is bi-hamiltonian and 
admits a Lax-pair formulation. It encompasses two different 
$N=2$ KdV hierarchies, the $a=4$ and $a=-2$ ones, which follow from it via 
appropriate reductions. Recently, one more integrable 
hierarchy with ``small'' $N=4$ SCA as the second hamiltonian structure has 
been found \cite{DGI}. As distinct from the ``genuine'' $N=4$ KdV system,  
$N=4$ global supersymmetry is broken to $N=2$ in this new ``quasi'' 
$N=4$ KdV hierarchy. It yields, by proper reductions, some previously 
unknown KdV type systems with lower supersymmetry. 

In this paper we extend the class of $N=4$ supersymmetric 
hierarchies by suggesting a general Lie-algebraic framework for them. 

Our consideration relies upon the well-known property that generalized KdV 
and KdV type hierarchies associated with (super)conformal algebras and 
$W$ algebras as the hamiltonian structure can be induced, via a 
Miura - type transformation, from the hierarchies 
associated in a similar way with appropriate (super)affine algebras. 
Thus the latter hierarchies underly the former ones, and are in a sense 
more fundamental.
In the algebraic language, such a correspondence 
amounts to the existence of the Sugawara-Feigin-Fuks (SFF) type (or coset) 
construction of the former (super)algebras in terms of the latter ones. For 
$N=2$ hierarchies a few examples of this sort were explicitly 
elaborated. These are, e.g., the relationships between $N=2$ KdV and 
$N=2$ Boussinesq hierarchies, on one hand, and their affine counterparts, 
on the other \cite{{DGI},{EISK}}. The relevant Miura transformations relate 
the corresponding second hamiltonian structures, $N=2$ SCA and $N=2$ $W_3$, 
to $N=2$ extensions of the affine algebra $\widehat{u(1)\oplus u(1)}$ 
and of two copies of the latter, respectively.   

Our proposal is to construct $N=4$ supersymmetric hierarchies 
in terms of $N=2$ extended affine algebras possessing a hidden 
$N=4$ structure. This class embraces the algebras admitting 
a quaternionic structure. Their local bosonic parts were listed 
in \cite{stp} while analyzing the question as to which group-manifold WZNW 
sigma models admit $N=4$ supersymmetric extension. The simplest 
non-trivial $N=2$ affine algebra of this kind is $N=2$ 
$\widehat{sl(2)\oplus u(1)}$, 
and this is the case we treat in detail in the present paper. 

We start with the formulation of $N=2$ $\widehat{sl(2)\oplus u(1)}$ 
in terms of two pairs of spin $1/2$ $N=2$ supercurrents subjected  
to the chirality-type constraints. We show that both the 
structure relations of the algebra and constraints are closed 
under some hidden nonlinear $N=2$ supersymmetry transformations. 
Together with the manifest 
$N=2$ supersymmetry, they constitute $N=4$ supersymmetry. 
The affine supercurrents form an irreducible multiplet of this $N=4$ SUSY. 
Then we demonstrate the existence of an infinite set of $N=4$ invariant 
quantities in involution constructed out of the affine supercurrents. 
This new hierarchy with $N=2$ $\widehat{sl(2)\oplus u(1)}$ 
as the second hamiltonian structure can be assigned the name 
$N=4$ super NLS-mKdV hierarchy because it leads to 
$N=2$ NLS hierarchy of ref. \cite{{ks},{kst}} and to
$N=2$ mKdV hierarchy as its two non-equivalent reductions. 
Further, we set up, via a generalized SFF consruction, 
three spin $1$ $N=2$ supercurrents of the ``small'' $N=4$ SCA. 
We demonstrate that in terms of the $N=4$ SCA
supercurrents the $N=4$ NLS-mKdV conserved quantities coincide 
with those of $N=4$ KdV hierarchy. Thus 
$N=4$ NLS-mKdV and KdV hierarchies are related to each other in the 
way resembling the relation between ordinary mKdV and KdV ones.
An essential difference is, however, that the affine algebra relevant 
to our case is the non-abelian  $N=2$  $\widehat{sl(2)\oplus u(1)}$
algebra. The ``quasi'' $N=4$ KdV hierarchy of ref. \cite{DGI} 
is also contained in 
the enveloping algebra of $N=2$ $\widehat{sl(2)\oplus u(1)}$. 
It corresponds to another choice of the evolution equations for 
the affine supercurents, with the hamiltonians breaking 
global $N=4$ supersymmetry down to $N=2$. 

\vspace{0.2cm}
\noindent{\bf 2. Hidden $N=4$ supersymmetry of $N=2$  
${\widehat{sl(2)\oplus u(1)}}$ algebra}.
The $N=2$ affine ${\widehat{sl(2)\oplus u(1)}}$ algebra  is constituted by 
the two pairs of $N=2$ spin $1/2$ fermionic superfields 
$F(Z)$, $\bar F(Z)$ and $H(Z)$, $\bar H(Z)$, $Z = (x,\theta, \bar \theta)$ 
being coordinates of $N=2$ superspace. These supercurrents satisfy 
the following classical Poisson Bracket (PB) relations 
\cite{hs}
\begin{eqnarray}
\{ H(1), {\overline H}(2) \} &=& D {\overline D} \delta (1,2)
\label{hhbr} \\
\{ H (1) , F(2) \} &=& D F \delta (1,2)\; , 
\;\;\;
\{ H (1) , {\overline  F}(2) \} =
-D {\overline F} \delta (1,2)\; , \nonumber\\
\{ {\overline H} (1) , F(2) \} 
&=&-{\overline D} F \delta (1,2) \; , \;\;\;
\{ {\overline H} (1) , {\overline F}(2) \} =
 {\overline D}  {\overline F}  \delta (1,2)\;, 
\label{u2al2}\\
\{F(1), {\overline F}(2) \}&=& 
\left[ (D + H)({\overline D} + {\overline H}) + 
F {\overline F} \right] \delta (1,2)\; ,
\label{u2al3}
\end{eqnarray}
all other PBs vanishing.
In these relations, $1,2 \equiv Z_{1,2}$, $D, {\overline D}$ 
are $N=2$ spinor derivatives 
\begin{eqnarray}
D = \frac{\partial}{\partial \theta}
-\frac{1}{2}{\overline \theta} \partial_x \;,\;\;
{\overline D} = \frac{\partial}{\partial 
{\overline \theta}} - \frac{1}{2}\theta\partial_x\; ,\;\;
D^2 = {\overline D}^2 = 0 \; , \;\; 
\{ D, {\overline D}\} = -\partial_x \; \label{Dcomm}\;, 
\end{eqnarray}
$\delta(1,2)$ is the $N=2$ superspace delta-function 
$$
\delta(1,2) = \delta(x_1 - x_2)(\theta_1 - \theta_2)
(\bar \theta_1 - \bar \theta_2)\;, 
$$
and the differential operators in the r.h.s. are evaluated at the point 
$Z_1$. The derivatives in the r.h.s. of these relations are assumed to act 
freely to the right. 

The PB set \p{hhbr} - \p{u2al3} is closed, i.e. 
the Jacobi identities are valid, only provided the involved supercurrents 
are subjected to the constraints \cite{hs} 
\begin{eqnarray}
D H &=& 0,\;\; \;\; \;{\overline D} {\overline H} = 0, \label{chir}\\
\left( D  + H\right)F &=& 0,\;\;\left( {\overline D}-
{\overline H} \right){\overline F} =0 \;. \label{covchir}
\end{eqnarray}
Thus $H$ and ${\overline H}$ are ordinary chiral and anti-chiral 
superfields while $F$ and ${\overline F}$ are ``covariantly'' 
chiral and anti-chiral ones.
It is easy to check that constraints \p{chir} - \p{covchir} are 
consistent with the above PB relations, in the sense that 
the PBs between them and with all supercurrents are vanishing on the shell 
of constraints. Note that the non-linear term in the r.h.s. of \p{u2al3} 
is ``fake'': at the component (and even at 
the $N=1$ superfield) level there is no non-linearity 
in the algebra. Its presence in \p{u2al3} is the price for 
manifest $N=2$ supersymmetry.

As is seen from \p{hhbr}, the superfields $H$ and ${\overline H}$ generate 
the Cartan $N=2$ $\widehat{u(1)\oplus u(1)}$ subalgebra of the  
$N=2$ $\widehat{sl(2)\oplus u(1)}$. The relations \p{u2al2} 
tell us about the properties of $F$ and ${\overline F}$ with respect 
to two independent $u(1)$ charges inherent in $N=2$ 
$\widehat{sl(2)\oplus u(1)}$. 
It is assumed that $H$ 
and ${\overline H}$ have the same transformation properties under the  
global $U(1)$ automorphism group of $N=2$, $1D$ superalgebra as the 
derivatives $D$ and ${\overline D}$.  

We note that for classical affine algebras
the central charge corresponding to the Kac-Moody level can be rescaled at
will (using the freedom to rescale the fields as well as the Poisson 
brackets). Our convention is dictated by further convenience.

The algebra $N=2$ $\widehat{sl(2)\oplus u(1)}$ is a particular 
representative of a wide class of $N=2$ affine algebras. Actually, 
$N=2$ superextension can be defined for any affine algebra (and superalgebra) 
admitting a complex structure \cite{hs}, \cite{ais}. 
$N=2$ supercovariance of these extensions is manifest 
when they are formulated in terms of (covariantly) chiral and anti-chiral 
$N=2$ superfields. Nevertheless, it could be equally revealed in the 
$N=1$ superfield formulation. In such a formulation, $N=2$ supercurrents 
are represented by pairs of real spin $1/2$ $N=1$ superfields, and 
rigid $N=2$ 
supersymmetry is realized as some transformations mixing $N=1$ superfields 
inside each pair. Then the statement that the given $N=1$ extended 
affine algebra is actually $N=2$ extended amounts to the covariance of 
the defining $N=1$ superfield PBs under these transformations. 

It is natural to ask whether the variety of $N=2$ affine 
(super)algebras contains as a subclass the algebras which reveal 
covariance under some extra supersymmetries. As should be clear 
from the reasoning just given, one way to answer this question is 
to explicitly construct the transformations of these additional 
supersymmetries on the affine $N=2$ supercurrents and to 
check covariance of the relevant PBs and chirality constraints 
with respect to them. 

Let us show that the algebra $N=2$ $\widehat{sl(2)\oplus u(1)}$ 
possesses a hidden $N=2$ supersymmetry which, together 
with the manifest $N=2$ supersymmetry, form $N=4$ 
supersymmetry. 

This extra supersymmetry is realized by the following non-linear 
transformations 
\begin{eqnarray}
\delta H &=& \epsilon D {\overline F} + {\overline \epsilon} H F,\;\;\;
\delta{\overline H} = {\overline \epsilon} {\overline D} F - \epsilon
{\overline H} {\overline F}\nonumber\\
\delta F &=& - \epsilon D {\overline H} - \epsilon ( H{\overline H}
+ F {\overline F}), \;\;\;
\delta {\overline F} = - {\overline \epsilon} {\overline D} H -
{\overline \epsilon } ( H {\overline H} + F {\overline F}),
\label{transform}
\end{eqnarray}
$\epsilon, {\overline \epsilon}$ being the corresponding infinitesimal 
parameters. 
It can be easily checked that the above transformations indeed realize 
an $N=2$ supersymmetry, i.e. that their commutators close to give, 
for any superfield $\Psi \equiv \{H, {\overline H}, F, {\overline F}\}$ 
\begin{eqnarray}
[ \delta_1, \delta_2 ] \Psi 
&=& ({\overline\epsilon}_1\epsilon_2 - {\overline\epsilon}_2\epsilon_1)
\partial_x \Psi\;.
\end{eqnarray}
The hidden and manifest $N=2$ supersymmetry transformations commute 
with each other and so form an $N=4$ 
supersymmetry. The two pairs of affine $N=2$ supercurrents 
$F, {\overline F}$ and $H, {\overline H}$ are unified into 
an irreducible $N=4$ supermultiplet. 

A remarkable property of the above transformations 
is that they preserve the (nonlinear) chirality and antichirality 
constraints \p{chir}, \p{covchir}. 

Moreover, it is a matter of tedious though straightforward 
computation to check that the defining relations of the 
$N=2$ $\widehat{sl(2)\oplus u(1)}$ algebra, eqs. \p{hhbr} - \p{u2al3} 
(together with the vanishing PBs), are closed under 
the transformations \p{transform}. One varies both sides of these 
relations and finds the variations being the same. 

Thus the algebra $N=2$ $\widehat{sl(2)\oplus u(1)}$ reveals a hidden 
$N=4$ supersymmetry and hence it is in fact a minimal $N=4$ extension of 
$\widehat{sl(2)\oplus u(1)}$. We expect that this $N=4$ covariance can be 
made manifest by combining the affine $N=2$ supercurrents 
into a single  $N=4$ superfield, e.g., in the framework 
of $N=4$ harmonic superspace \cite{DI}.

It is reasonable to assume that there exists the whole class of affine 
algebras and superalgebras allowing for an $N=4$ extension along simialr 
lines. A plausible conjecture is that these algebras are those which 
admit a quaternionic structure. The full list of such algebras and 
corresponding groups was given 
in \cite{stp}.  
Though we are not still aware of the general proof of this conjecture, 
we have checked it on more examples. One of them is the $N=2$ affine 
$\widehat{sl(3)}$ algebra. Its basic relations also reveal 
covariance under the appropriate $N=4$ transformations, we are going 
to consider this case in detail in 
a forthcoming publication. Another, simplest example is 
an $N=2$ extension of the affine algebra 
${\oplus_{i=1}^{4}} \widehat{u(1)_i}$. It is constituted by the two 
conjugated mutually commuting pairs of chiral and anti-chiral 
$N=2$ superfields $H_\alpha, {\overline H}_\alpha$, $(\alpha=1,2)$, 
\bea
\{ H_\alpha, H_\beta \} &=& 0\;, \;\;\;\;  \{ H_\alpha,  
{\overline H}_\beta \} = 
\delta_{\alpha\beta} D{\overline D} \delta(1,2)\;, \label{u111} \\
DH_\alpha &=& {\overline D} {\overline H}_\alpha = 0 \;. \nonumber 
\eea
It is straightforward to check the covariance of \p{u111} 
under the linearized version of \p{transform}
\begin{eqnarray}
\delta H_1 = \epsilon D {\overline H}_2, \;\; 
\delta{\overline H} = {\overline \epsilon} {\overline D} H_2, \;\; 
\delta H_2 = - \epsilon D {\overline H}_1,\;\; 
\delta {\overline H}_2 = - {\overline \epsilon} {\overline D} H_1\;. 
\label{transform2}
\end{eqnarray}
These transformations has the same Lie bracket structure as \p{transform}.
Later on we will argue that this simplest $N=2$ affine algebra with 
$N=4$ structure bears no direct relation to $N=4$ KdV hierarchy of refs. 
[4-6], in contrast to $N=2$ $\widehat{sl(2)\times u(1)}$ which does.

\vspace{0.2cm}
\noindent
{\bf 3. $N=4$ invariant hamiltonians and flows}.
In the previous section we have shown that the algebra 
$N=2$ ${\widehat{sl(2)\oplus u(1)}}$ 
carries an $N=4$ structure and is in fact an $N=4$ extension of 
${\widehat{sl(2)\oplus u(1)}}$. We may wonder whether it 
generates, as a second hamiltonian structure, some $N=4$ supersymmetric 
hierarchy of evolution equations for $F, {\overline F}, H, {\overline H}$, 
or, in other words, whether its enveloping algebra contains an infinite 
set of $N=4$ supersymmetric hamiltonians in involution. The answer is 
positive, and now we explicitly furnish first three such $N=4$ invariant 
hamiltonians. The proof that we indeed face an 
integrable hierarchy will be given later by presenting the scalar Lax 
operator for it.

The hamiltonian densities ${\cal H}_i$ given below are
invariant under (\ref{transform}) up to total derivatives
and correspond to 
integral spin dimension
$i=1,2,3$, respectively. In addition, we require
them to be globally 
chargeless ($Q({\cal H}_i) = 0$) with respect to a charge operator $Q$
such that $Q(H) = Q({\overline H})=0$, $Q(F)= 1$, $Q({\overline F}) = -1$.
The reason why we impose this condition is the desire to reproduce 
the conserved quantities of $N=2$ NLS hierarchy in the limit 
$H={\overline H} = 0$; requiring this invariance amounts to the property 
that the reduced hamiltonians ``live'' on the quotient of $N=2$
$\widehat{sl(2)\oplus u(1)}$ over its Cartan subalgebra 
$\widehat{u(1)\oplus u(1)}$, which is the cahracteristic feature of 
$N=2$ NLS hierarchy \cite{kst}.
For sure, this choice does not yield the most general set of $N=4$ 
invariant hamiltonians one can construct 
(and does not even correspond to the most general hierarchy, see
however the remarks in section 4), but it is the only choice
which allows us to recover $N=2$ NLS hierarchy.

It is of interest to point out that ${\cal H}_2$ is uniquely 
defined by the requirements of $Q({\cal H}_2)=0$ and $N=4$ invariance, 
i.e. it is of no need to resort to the involutivity 
reasonings to specify its coefficients. We have
\begin{eqnarray}
{\cal H}_1 &=& F {\overline F} + H {\overline H}\nonumber\\
{\cal H}_2 &=& F' {\overline F} - H' {\overline H}  -
(D{\overline H} + {\overline D} H) (H{\overline H} + F {\overline F})
- 2 H {\overline H} F {\overline F}\nonumber\\
{\cal H}_3 &=& F'' {\overline F} + H'' {\overline H} - 
  3{\overline D}HF{\overline F}'-6{\overline D}HF'{\overline F}'+
  3{\overline D}HH'{\overline H}+ 
  6D{\overline H}F{\overline F}'-H{\overline D}F{\overline F}'\nonumber \\
 & & -\; 5H'{\overline D}F{\overline F}+\frac{3}{2}H'D{\overline H}
{\overline H}-
    4 {\overline H}F'D{\overline F}-   {\overline H}'FD{\overline F}-
  {\overline D}FFD{\overline F}{\overline F}\nonumber \\
 & & + \; 9{\overline D}H{\overline D}HF{\overline F}+
  {\overline D}H{\overline D}HH{\overline H}-
  8{\overline D}HH{\overline D}F{\overline F}+ 
   3{\overline D}HHD{\overline H}{\overline H} \nonumber \\
 & &+ \; 9D{\overline H}D{\overline H}F{\overline F}+
   8D{\overline H}{\overline H}FD{\overline F}+
  HD{\overline H}D{\overline H}{\overline H}+
  2H{\overline H}{\overline D}FD{\overline F}   \label{3dens}
\end{eqnarray}
(hereafter, the spatial derivative is denoted with a prime).

The corresponding first and second flows are given by
\begin{eqnarray}
{\partial  \Psi \over \partial t_1} & =& \Psi '
\end{eqnarray}
for any $\Psi \equiv\{H, {\overline H}, F, {\overline F}\}$, and
\begin{eqnarray}
{\partial F\over \partial t_2} &=&  
- F'' 
+ 2{\overline D} H F'
+ 2 {\overline D}H' F 
+ 2 D{\overline H} F'
+ 2 {\overline D} F F D{\overline F} - 2 F F' {\overline F} \nonumber \\
&& +\; 2 H D{\overline H} {\overline D} F 
+ 2 H {\overline H} F' 
+ 2 H {\overline H}' F - 2 {\overline D}H H {\overline H} F +
4 H {\overline D} F F {\overline F}\;,   \nonumber\\
{\partial H \over \partial t_2} &=&  
H'' 
+ 2 {\overline D} H H' 
+ 2 H D{\overline H}' 
+ 2 H'D{\overline H} + 2 {\overline D} H F D{\overline F} 
- 2 H {\overline D} F D {\overline F} \nonumber \\
&& + \; 2 H F' {\overline F} +
2 H' F {\overline F} - 2 H D{\overline H} F {\overline F}
- 2 H {\overline H} F D {\overline F} \;. \label{eof}
\end{eqnarray}
We do not present the third flow equations in view of their complexity. 

\vspace{0.2cm}
\noindent
{\bf 4. Generalized Sugawara construction and relation to $N=4$ KdV}.
To clarify the meaning of the above involutive sequence of $N=4$ 
supersymmetric hamiltonians, let us apply to the Sugawara type construction 
on the algebra $N=2$, $\widehat{sl(2)\oplus u(1)}$. We will show 
that it naturally leads to the classical version of ``small'' $N=4$ 
SCA with the arbitrary central charge (proportional to that of 
the affine algebra). In terms of composite $N=4$ SCA supercurrents 
the hierarchy presented in the previous secton is recognized as 
$N=4$ KdV hierarchy. 

We start with the standard classical $N=2$ SFF 
stress-tensor (with a fixed central charge, see remark in Sect. 2)
\be
J = H{\overline H} + F {\overline F} + D {\overline H} 
+ {\overline D} H \;. \label{N2stress}
\ee
As a consequence of the $N=2$ affine PBs \p{hhbr} - \p{u2al3} it 
satisfies the PB of $N=2$ SCA 
\be 
\{ J(1), J(2) \} = \left( J\partial+ \partial J  + DJ {\overline D} + 
{\overline D}J D + \partial [D,{\overline D}] \right) \delta(1,2)\;, 
\label{N2pb}
\ee
where as before the differential operator in the r.h.s. is evaluated 
at $Z_1$ and all derivatives act freely. Considering the transformation 
properties of $J$ under the transformations \p{transform}, we find that 
it forms an irreducible {\it linear} $N=4$ supermultiplet together with 
the spin $1$ chiral and anti-chiral supercurrents 
\be  \label{defphi}
\Phi \equiv D{\overline F}\;,\;\;\; {\overline \Phi} \equiv 
{\overline D}F\;, \;\; D \Phi =  {\overline D} {\overline \Phi} = 0\;.
\ee
\be
\delta J = -\epsilon {\overline D} \Phi -\bar \epsilon D {\overline \Phi}\;, 
\;\; \delta \Phi = \bar \epsilon D J\;, \;\; \delta {\overline \Phi} = 
\epsilon {\overline D} J\;.
\label{transform3}
\ee
It is straightforward to check that the original PBs \p{hhbr} - \p{u2al3} 
imply the following non-vanishing PBs for $\Phi$, ${\overline \Phi}$
\bea 
\{ J(1), \Phi(2) \} &=& -\left(\Phi {\overline D}D + 
{\overline D}\Phi D \right) \delta(1,2)\;,
\{ J(1), {\overline \Phi}(2) \} = -
\left( {\overline \Phi} D{\overline D} + 
D{\overline \Phi}{\overline D} \right) \delta(1,2)\; ,
\nonumber \\
\{ \Phi(1), {\overline \Phi}(2) \} &=& \left(\partial D{\overline D} 
+DJ{\overline D} \right)\delta(1,2)\;. \label{N4pb}
\eea
Together with \p{N2pb} they form a closed set of PBs which defines  
a classical version of ``small'' $N=4$ SCA [5,8] (it is closed 
with respect to both Jacobi identities and rigid $N=4$ transformations 
\p{transform3}) \footnote{The precise correspondence with, e.g., 
ref. \cite{DGI} is achieved 
by rescalings $J \rightarrow -2J, \partial \rightarrow -\partial, 
\{\;, \;\} \rightarrow {1\over 2} \{\;, \;\} $.}.

As was remarked previously, in 
the present case it is not essential which specific value is ascribed 
to the central charge, since for classical algebras it 
can be rescaled to any value through a rescaling of the 
fields and PBs. However, it is worth mentioning that the central 
charge of the above $N=4$ SCA is strictly related to 
that of the underlying $N=2$ $\widehat{sl(2)\oplus u(1)}$ algebra. 
In particular, with our convention on the central charge in 
the structure relations of 
$N=2$ $\widehat{sl(2)\oplus u(1)}$ the coefficient before the 
Feigin-Fuks term in \p{N2stress} should 
be strictly $1$ in order the PBs of $N=4$ SCA were closed. 
We also note that now the rigid supersymmetry 
\p{transform} can be recovered as a part of local supersymmetry 
generated by  $\Phi$, ${\overline \Phi}$ on the affine $N=2$ supercurrents 
through the PBs \p{hhbr} - \p{u2al3}.

The $N=4$ SFF realization \p{N2stress}, \p{defphi} is the 
key which allows us to establish a connection between our $N=4$ 
NLS-mKdV hierarchy and the $N=4$ KdV. It is a simple exercise to 
check that the first three hamiltonian densities in \p{3dens}, up to 
full derivatives, are expressed in terms of the above composite $N=4$ 
supercurrents 
\begin{eqnarray}
{\cal H}_1 = J\;,\;\;
{\cal H}_2 = -{\textstyle {1\over 2}} \left( J^2 - 
2\Phi{\overline \Phi} \right)\;, \;\;
{\cal H}_3 = {\textstyle{1\over 2}} \left( J [D, {\overline D}] J 
+2 \Phi {\overline \Phi}'+ {\textstyle{2\over 3}} J^3
-4 J \Phi {\overline \Phi} \right)\;.  \label{N4expr}
\end{eqnarray}
They coincide (up to rescalings of supercurrents and conventions on 
the algebra of $N=2$ spinor derivatives) with the explicit expressions 
for the first three hamiltonians of $N=4$ KdV hierarchy given in 
\cite{DIK}.
 
For completeness let us write here the corresponding flows of the $N=4$
KdV hierarchy. Besides the first flow ($\dot{\Omega} = \Omega'$,  
$\Omega \equiv \{J, \Phi , {\overline \Phi} \}$) we have
\begin{equation}
{\partial \over \partial t_2} J = \left[D,{\overline D}\right] J' + 2 J J' 
  -2(\Phi{\overline \Phi})' \; ,
{\partial \over\partial t_2 } \Phi = \Phi'' - 
       2 D {\overline D} (J \Phi) \; .
\end{equation}

Our hamiltonians correspond to a particular choice ($a=4$, 
$b =0$ in \cite{DIK}), for the
$N=4$ integrable hierarchy involving $J, \Phi, {\overline \Phi}$. 
It has
been proven in \cite{DIK} that there exist 
at most two $N=4$ integrable hierarchies   
characterized by different choices of the coefficients. However the two
hierarchies so characterized are related to each other through a global 
transformation of the $N=4$ supersymmetry automorphism 
group $SU(2)$. This means that they actually define the
same $N=4$ hierarchy. This point has been fully analyzed in \cite{DIK},
both with the use of harmonic superspace and in the $N=2$ 
formalism. We limit ourselves to
pointing out that in the language of $N=2$ affine superfields 
the second choice of the coefficients would correspond to 
hamiltonians which do not obey the $U(1)$ chargeless condition 
$Q({\cal H})=0$.
As follows from the consideration in ref. \cite{DIK}, 
they still should respect some hidden $U(1)$ invariance. Their precise 
relation to the above hamiltonians with manifest $U(1)$ invariance 
can be revealed by finding out how the automorphism group $SU(2)$ is 
realized on $H, {\overline H}, F, {\overline F}$ and then performing 
an appropriate $SU(2)$ rotation of these hamiltonians.

It should be
also stressed that we cannot exclude the possibility that other $N=4$ 
hierarchies, involving the superfields $H, {\overline H}, F, {\overline F}$
but not expressible only through $J, \Phi, {\overline \Phi}$, 
could indeed exist. We leave the complete analysis of this 
problem for the future work.

We end this section with two comments. 

The first one regards a hidden $N=4$ 
structure of the $N=2$ $\widehat{sl(2)\oplus u(1)}$ algebra. It turns out 
that the latter gives rise, besides the ``small'' $N=4$ SCA, 
to a wider $N=4$ SCA, namely to the 
``large'' $N=4$ SCA  in a specific realization. Indeed, one 
can define one more $N=4$ supermultiplet of composite supercurrents 
\bea 
\hat{J} &=& H{\overline H} + F{\overline F}\;, \;\; 
\hat{\Phi} = HF\;, \;\; \hat{{\overline \Phi}} = 
{\overline F}{\overline H}\;, \label{newcomp} \\
\delta \hat{J}  &=& \bar \epsilon {\overline D} \hat{\Phi} - \epsilon 
D \hat{\overline \Phi}\;,\;\; \delta \hat{\Phi} = -\epsilon D \hat{J}\;,\;\; 
\delta \hat{\overline \Phi} = -\bar \epsilon {\overline D}\hat{J}\;, \nonumber
\eea
which satisfy the PBs \p{N4pb}, but with vanishing central terms. 
In other words, 
they generate a kind of topological ``small'' $N=4$ SCA. Together with 
$J$, $\Phi$, ${\overline \Phi}$ these additional supercurrents close 
on ``large'' $N=4$ SCA in a particular realization with 3 elementary 
bosonic $sl(2)$ spin $1$ affine currents and three composite ones 
constructed out of the elementary fermionic spin $1/2$ currents. 
The presence of this hidden ``large'' $N=4$ SCA in $N=2$ affine 
$\widehat{sl(2)\oplus u(1)}$ algebra has been earlier noticed, at 
the full quantum level, in \cite{KKK}. This feature can be considered as 
an indication that 
the presented $N=4$ NLS-mKdV hierarchy bears a relation to 
a more general super KdV hierachy (still to be constructed) associated 
with the whole ``large'' $N=4$ SCA.

The second comment is related to $N=2$ extension of 
the abelian affine algebra $\oplus_{i=1}^4 \widehat{u(1)}_i$, 
eqs. \p{u111}. As was noticed in Sect. 2, it also reveals 
a covariance under the rigid 
$N=4$ supersymmetry, this time realized by linear transformations 
\p{transform2} (combined with those of manifest $N=2$ supersymmetry). 
One may 
wonder whether it also gives rise to $N=4$ SCAs via some kind of Sugawara 
construction and has any relation to $N=4$ KdV hierarchy. It is easy 
to check that one cannot construct, out of the superfields 
$H_\alpha, {\overline H}_\beta$, any $N=4$ multiplet of composite 
currents which would include $N=2$ conformal stress-tensor with 
a Feigin-Fuks term (the latter is absolutely necessary for 
producing a central term in $N=2$ SCA and thus generating at least 
an $N=2$ KdV hierarchy). The only possibility is the $N=4$ multiplet 
\be
\hat{J} = H_1{\overline H}_1 + H_2{\overline H}_2,\;\;\; \hat{\Phi} 
= H_1H_2, \;\; \hat{{\overline \Phi}} = {\overline H}_2 {\overline H}_1\;, 
\ee
which, via PBs \p{u111}, generates a topological ``small'' 
$N=4$ SCA. So, possible $N=2$ hierarchies (even posessing 
rigid $N=4$ supersymmetry) constructed on the basis 
of $N=2$ $\oplus_{i=1}^{4} \widehat{u(1)}_i$ algebra seem to have 
no any direct relation to $N=4$ KdV 
hierarchy. We have also explicitly checked that there exists no  
$N=4$ covariant system of evolution equations for 
$H_\alpha, {\overline H}_\beta$ which would reduce to $N=2$ NLS 
system of refs. \cite{{ks},{kst}} after putting one pair of these superfields 
equal to zero. Thus the minimal way to define an mKdV type hierarchy for  
$N=4$ KdV system (yielding also $N=2$ NLS hierarchy as a 
consistent reduction) is based on the use of non-abelian affine superalgebra 
$N=2$ $\widehat{sl(2)\oplus u(1)}$. The nonlinearity of rigid $N=4$ 
transformations \p{transform} is crucial for constructing the 
$N=4$ multiplet 
of supercurrents \p{N2stress}, \p{defphi} with a non-trivial 
$N=2$ SFF stress-tensor.

\vspace{0.2cm} 
\noindent
{\bf 5. Lax operator}.
Due to the Sugawara-Feigin-Fuks construction which maps the 
affine superfields $H, {\overline H}, F, {\overline F}$ onto the 
superfields generating the $N=4$ SCA there is no further need to prove 
the integrability of the $N=4$ NLS-mKdV hierarchy. 
We are indeed 
guaranteed by the integrability property of the $N=4$ KdV hierarchy 
that the corresponding 
affine hierarchy is integrable. The higher order hamiltonians in 
involution are in fact expressed through $J, W, {\overline W}$ and 
therefore, via SFF, through $H, {\overline H}, F, {\overline F}$.

It is, however, important to realize that the 
Lax operator for the $N=4$ KdV \cite{DG}
indeed furnishes the correct evolution equations for the affine superfields,
once the basic superfields are re-expressed by the SFF construction.
This statement has to be verified
independently because, to our knowledge, no general realtionship was 
as yet established between the $N=2$ affine PB structure and the Lax 
operators we are considering. So we made this checking explicitly.

The $N=4$ NLS-mKdV scalar Lax operator constructed 
according to the above prescription reads
\begin{equation}  \label{lax1}
L= D{\overline D} + D {\overline D} \partial^{-1} 
\left( J + {\overline  \Phi} \partial^{-1} \Phi \right) 
\partial^{-1} D{\overline D}\;,
\end{equation}
where $J$ and $\Phi, {\overline \Phi}$ are expressed by 
eqs. \p{N2stress}, \p{defphi}.
The flows are given by
\begin{eqnarray}  \label{lax1eq}
\frac{\partial}{\partial t_k} L &=& - [ {L^{k}}_{\geq 1}, L ]
\end{eqnarray}
where the suffix $\geq 1$ means taking the strictly differential part of
the operator.

\vspace{0.2cm}
\noindent{\bf 6. $N=2$ reductions and bosonic cores}.
In this section we discuss different $N=2$ supersymmetric integrable 
reductions of the $N=4$ NLS-mKdV hierarchy.

\vspace{0.2cm}
\noindent{\it i)} Let us firstly set $F= {\overline F} = 0$. 
This is a  consistent 
reduction for all flows as all the time derivatives of $F$ are zero 
in this limit. 
The hierarchy thus produced 
is expressed in terms of the chiral and antichiral superfields 
$H, {\overline H}$ generating the ${\widehat{u(1)\oplus u(1)}}$ 
subalgebra as the relevant second hamiltonian structure. 
It is the $N=2$ mKdV hierarchy associated with the $a=4$, $N=2$ 
KdV system. 
Indeed, upon this reduction, $\Phi={\overline \Phi} =0$, 
$J= (H {\overline H} + D {\overline H} +{\overline D} H)$, and 
the closed set of the evolution equations for $J$ 
corresponds just to the $a=4$, $N=2$ KdV. This reduction can be performed  
directly in the Lax representation \p{lax1}, \p{lax1eq}. 

\vspace{0.2cm}
\noindent{\it ii)} Let us set $H={\overline H} = 0$. Then 
\be
J = F {\overline F}, \;\; \Phi = D{\overline F}, \;\; {\overline \Phi} 
= {\overline D}F\;,\;\;  DF = {\overline D} {\overline F} = 0\;.
\label{red2}
\ee
This is also a consistent reduction 
because both the left- and right-hand 
sides of the evolution equations for $H={\overline H} = 0$ vanish 
upon effecting it. It cannot be directly performed 
at the algebraic level, we have to implement the Dirac's bracket formalism 
to deduce the relevant second hamiltonian structure algebra
\footnote{This non-local structure is explicitly quoted in \cite{shurik1}.}. Nevertheless, 
it goes straightforwardly at the level of the equations and 
conserved hamiltonians and yields the $N=2$ NLS hierarchy of 
refs. \cite{{ks},{kst}}. It is a simple exercise to check that the 
conserved quantities $H_1$ - $H_4$ of $N=4$ NLS-mKdV (or $N=4$ KdV) 
hierarchy go over to the corresponding quantities of $N=2$ NLS one 
(with $H = {\overline H} = 0$) upon substituting the expressions 
\p{red2} for $J$ and $\Phi$, ${\overline \Phi}$. The fact that the 
reduction \p{red2} takes the Lax operator for $N=4$ KdV hierarchy 
into that for $N=2$ NLS has been earlier noticed in \cite{IK}.

\vspace{0.2cm}
\noindent{\it iii)} There exists one more, rather unexpected reduction. 
It goes by imposing a sort of ``mixed'' constraint, ${\overline H}=0$ 
and $F=0$ (or equivalently $H=0$ and ${\overline F}=0$). 
The constraint $F=0$ is allowed due to the global charge conservation, 
but it turns out that also ${\overline H} =0$ is admitted by 
the equations. As in the previous case, these two constraints can be 
straightforwardly imposed at the level of the evolution equations  
while requiring a more subtle treating at 
the level of PBs. For completeness we give here
the flows corresponding to this ``mixed'' reduced $N=2$ hierarchy. 
It would be interesting to analyze its possible relation to
other known $N=2$ hierarchies. Besides the trivial first flow we have
\begin{eqnarray}
{\partial H \over \partial t_2} = H'' + 2 {\overline D} H H'\;,\;\; \;
{\partial  {\overline F}\over\partial t_2} = 
{\overline F}'' + 2 {\overline D} H {\overline F}'\;, 
\end{eqnarray} 
and
\begin{eqnarray}
{\partial H \over \partial t_3} &=& - H''' - 3 H''{\overline D}H -
3H' {\overline D} H {\overline D} H - 3 {\overline D} H' H'\nonumber\\
{\partial {\overline F}\over \partial t_3}  &=&  - {\overline F}''' - 3 {\overline F}''
{\overline D} H - 3 {\overline F}' {\overline D H} {\overline D} H
- 3 {\overline F}' {\overline D}H' \;.
\end{eqnarray}
Thus in both cases we get two non-linear equations, 
one involving the superfield $H$ alone and the other 
describing an evolution of an anti-chiral superfield ${\overline F}$ 
in the background of $H$.

Finally, let us present the bosonic limit of the second flow equations of 
our $N=4$ mKdV-NLS hierarchy. 
We define the bosonic components of the superfields 
$H,{\overline H},F,{\overline F}$ as follows
\begin{equation}
{\overline D}F|=\phi, D{\overline F}|={\overline\phi},
{\overline D}H|=h, D{\overline H}|={\overline h}.
\end{equation}
With such a definition the bosonic core of our system \p{eof} reads
\begin{eqnarray}
\frac{\partial\phi}{\partial t_2} & = & -\phi''+2h\phi'+
  2h'\phi+2{\overline h}\phi'+
  2\phi\phi{\overline \phi}+2h{\overline h}{\overline \phi} ,
   \nonumber \\
\frac{\partial h}{\partial t_2} & = & h''+2hh'+
  2h{\overline h}'+ 2h'{\overline h}.
\end{eqnarray}

\vspace{0.2cm}
\noindent{\bf 7. One more $N=2$ hierarchy with the 
$\widehat{sl(2)\oplus u(1)}$ structure}.
In {\cite{DGI} it has been proven that ``small'' $N=4$ SCA 
provides the PB structure for one more hierarchy of integrable
equations whose hamiltonians respect only $N=2$ supersymmetry. 
It is an extension of the $a=-2$, $N=2$ KdV hierarchy and 
is related via a non-polynomial
Miura transformation to the $\alpha = -2$, $N=2$ super Boussinesq 
hierarchy (and hence to $N=2$ ${\cal W}_3$ superalgebra which is the 
second PB structure for all three known $N=2$ Boussinesq hierarchies). 
Its characteristic feature is that all the
even-dimensional hamiltonians are vanishing in the $a=-2$, $N=2$ 
KdV limit $\Phi = {\overline \Phi} = 0$. 
First non-trivial hamiltonian is given by the following integral
over $N=2$ superspace
\begin{eqnarray}  
H_2 & = & \int dZ \Phi {\overline \Phi} = 
\int dZ D{\overline F}\; {\overline D}F.   \label{N2conserv}
\end{eqnarray}

It is clear that the existence of this hierarchy implies, via the 
$N=4$ SFF construction \p{N2stress}, \p{defphi}, 
the existence 
of the underlying mKdV type heirarchy of evolution equations for 
the $N=2$ affine superfields $H, {\overline H}, F, {\overline F}$. 
Indeed, substituting into the relevant hamiltonians the expressions  
\p{N2stress}, \p{defphi} for $J$ and $\Phi, {\overline \Phi}$ 
one easily deduces 
the equations for $H, {\overline H}, F, {\overline F}$ using 
the PBs algebra \p{hhbr} - \p{u2al3}. We limit ourseleves by 
explicitly presenting the second flow equations associated with $H_2$, 
eq. \p{N2conserv}
\begin{eqnarray}
 {\partial F\over \partial t_2} &=&
-F'' + F' F {\overline F}
-H' {\overline D} F
+ D{\overline H} F'
+{\overline D} H' F
+{\overline D} H F'-{\overline D} H D{\overline H} F \nonumber \\
&&+ \;H D{\overline H} {\overline D} F
- H {\overline D} H {\overline H} F
+ H {\overline H} F'
+ H F {\overline D} F {\overline F}\;, \nonumber\\
{\partial H\over \partial t_2} &=&
- F D {\overline F}' - F' D{\overline F}
- H F {\overline F}'
- 
H {\overline D} F D{\overline F}
+ {\overline D} H F D{\overline F} 
\nonumber \\
&& - \; H {\overline H} F D{\overline F}
-H D{\overline H} F {\overline F}
\;.
\end{eqnarray}

To all consistent reductions of the ``quasi'' $N=4$ KdV hierarchy 
listed in \cite{DGI} there correspond the appropriate reductions 
of this underlying mKdV type hierarchy. In particular, the 
reduction $F={\overline F} = 0$ is consistent, and it takes the above  
hierarchy into the $N=2$ mKdV one corresponding to the $a=-2$, $N=2$ KdV 
hierarchy. Upon this reduction all the even-dimension hamiltonians vanish 
and the corresponding flows become trivial. Note that  
no reduction $H={\overline H} = 0$ exists in the present case , which 
reflects the fact that no analog of $N=2$ NLS hierarchy can be 
defined for the $a=-2$, $N=2$ KdV.

Like in the previous case, scalar Lax representation for the above hierarchy 
can be obtained by substituting \p{N2stress}, \p{defphi} into 
the ``quasi'' $N=4$ KdV heirarchy Lax operator given in \cite{{DGI},{DG}}
\begin{eqnarray}
L &= & D\left( \partial +H{\overline H}+F{\overline F} +
({\overline D}H)+(D{\overline H})-(D{\overline F})\partial^{-1}
   ({\overline D}F)\right){\overline D} , \\
\frac{\partial}{\partial t_k}L & = & -\left[ L^{\frac{k}{2}}_{\geq 1},L
 \right] \;. \nonumber
\end{eqnarray}

\vspace{0.2cm}
\noindent{\bf 8. Conclusions}.
In this paper we have succeeded for the first time in establishing 
a link between $N=2$ affine Lie algebras on one hand and 
large $N$ supersymmetric extended hierarchies on the other. 
We did so by showing that the $N=2$ 
affine $\widehat{sl(2)\oplus u(1)}$ algebra reveals a global 
$N=4$ supersymmetry and yields a PB structure for a globally $N=4$ 
supersymmetric integrable
hierarchy which generalizes both $N=2$ NLS and $N=2$ mKdV ones. 
We proved that an appropriate Sugawara-Feigin-Fuks construction 
leads to the ``small'' $N=4$ SCA and
to the related $N=4$ KdV. As a by-product of our method some other 
constructions have been worked out. In particular we have furnished a new 
global $N=2$ hierarchy based on the same affine algebra. Lax operators 
have been given for both these hierarchies.

It is clear that this work opens a way to further developments. Just
as in the purely bosonic case, where the constructions based on affine Lie
algebras have paved the way towards an understanding and a  
classification of all hierarchies, a similar situation can now be faced
for the $N=4$ hierarchies. The strategy is clear, it should consist 
in selecting affine algebras with quaternionic structure and 
analyzing the properties of their
affinizations, sequences of invariant hamiltonians in involutions, 
Sugawara constructions, etc. The next simplest case of interest 
involves the $N=2$ superaffine $\widehat{sl(3)}$ algebra, 
which one naturally suspects to lead to $N=4$ generalizations 
of (some of) the $N=2$ Boussinesq hierarchies and to a sort of 
$N=4$ extensions of $W_3$ algebra.

All things could probably be made much more simple and 
transparent in a manifestly $N=4$ supersymmetric formalism, 
based, e.g., on the harmonic superspace approach.

Finally, we note that in a recent paper \cite{shurik2}  
$N=4$ KdV hierarchy was mapped on the so called $(1,1)$ 
GNLS hierarchy \cite{bonora} defined on a pair of chiral and anti-chiral 
fermionic and bosonic superfields. It is unclear to us which algebraic 
structure underlies this non-polynomial map and whether it bears 
any relation to the construction presented here.

\vskip1cm
\noindent{\Large{\bf Acknowledgments}} \\
E.A. and S.K. acknowledge a partial support from the 
RFBR-DFG project No 9602-00180. Their work was also partly 
supported by RFBR under the project No 96-02-17634, INTAS under the 
project INTAS-94-2317 and by a grant of the Dutch NWO organization.
F.T. would like to thank the Director of the Bogoliubov 
Laboratory of Theoretical Physics (JINR, Dubna) Academician 
D.V. Shirkov for the hospitality at BLTP during the course of this 
work. S.K. would like to thank D. L\"{u}st for the hospitality at 
Humboldt University (Berlin) on the early stage of this study.

\end{document}